# AI pipeline for accurate retinal layer segmentation using OCT 3D images.

Mayank Goswami
Divyadrishti Imaging Laboratory, IIT Roorkee,
Indian Institute of Technology Roorkee, Roorkee, Uttarakhand, India, 247667;
mayank.goswami@ph.iitr.ac.in

Abstract:
Image data set from a multi-spectral animal imaging system is used to address two issues: (a) registering the oscillation in optical coherence tomography (OCT) images due to mouse eye movement and (b) suppressing the shadow region under the thick vessels/structures. Several classical and AI-based algorithms in combination are tested for each task to see their compatibility with data from the combined animal imaging system. Hybridization of AI with optical flow followed by Homography transformation is shown to be working (correlation value>0.7) for registration. Resnet50 backbone is shown to be working better than the famous U-net model for shadow region detection with a loss value of 0.9. A simple-to-implement analytical equation is shown to be working for brightness manipulation with a 1% increment in mean pixel values and a 77% decrease in the number of zeros. The proposed equation allows formulating a constraint optimization problem using a controlling factor α for minimization of number of zeros, standard deviation of pixel value and maximizing the mean pixel value. For Layer segmentation, the standard U-net model is used. The AI-Pipeline consists of CNN, Optical flow, RCNN, pixel manipulation model, and U-net models in sequence. The thickness estimation process has a 6% error as compared to manual annotated standard data.

## 1. Introduction

Optical coherence tomography (OCT) imaging system is a preferred imaging technique for depth and time-resolved 3D ocular imaging [1]. The technique provides time-dependent topographical structures of the deep retina in micrometre order. Besides ocular imaging, it is also used to study cardiovascular, dermatological lesions, and sub-surface cerebral activities [2–4]. Phase variance of OCT data can be correlated with dynamic structure thus providing a blood vessel map.

OCT as a functional imaging tool is used as a clinical diagnostic tool. It is also used in translational research in laboratory environments. Both, translational and clinical applications require quantitative analysis to compare the baseline images with respect to time or any other variable. Progression and prognosis are correlated with changes in thickness, density, color, and volume in search of biomarkers using multi-modal/multi-spectral imaging systems [5–7]. The thickness of layers can be used as one of the imaging biomarkers to differentiate between the retina of a healthy person/animal and the retina of a patient/animal suffering from a disease. However, for statistically sound data and strong disease correlation temporal resolution must be high. Ethically, once the patient is diagnosed with a disease, clinical guidelines and professional etiquette imposes upon the urgency to provide treatment immediately. Such situations do not allow for securing images for several days. Similar murine disease model (known and experimental both) imaging in laboratory setup is more often used as a surrogate that may facilitate insight specially to develop and test new pre-clinical treatment protocols [7,8].

OCT brightness scan (B Scan) images can be used to quantitatively estimate the thickness of the whole eye with better accuracy non-invasively in-vivo comparable to gold standard histopathological images [9]. However, estimation is always subjected to accurate image processing steps for example retinal segmentation.

### 1.1 Issue requiring post-processing manipulations.

#### 1.1.1. Oscillation due to mouse eye movement:

Usually, a mouse is put under anaesthesia for sake of convenience during the imaging. Animals, undergone several imaging sessions may develop resistance against optimal dosage. The dynamic adjustment (if the animal is made to inhale the isoflurane) is possible. However, if done during the imaging session, one may observe slight oscillations in B Scans. The retinal layers having different base heights/levels from the top of the B Scan are shown in figs. 1(a) to 1(c) using curly brackets and blue color double-sided arrow makers.

Multiple B Scans images of the same locations are repeatedly measured and averaged later during the post-processing step [10]. It provides images with a relatively better signal-to-noise (SNR) ratio. In our case, 1080 OCT B Scans are averaged into 360 using an adjacent group of 3. Figures 1(a), 1(b), and 1(c) show the SNR values of three repeated B Scans and their respective averaged B scans. Each figure shows zoomed Region of Interest depicting the presence of external limiting membrane (ELM) and retinal epithelial pigment (RPE) layers. Figure 1(d) shows multiple ELM and RPEs manifested as averaging is done without registration as the distance between the upper reference level to ILM increased (H2>H1), highest in fig. 1(c). An inaccurate registration may also create observable overlapped layers or the presence of fake retinal layers after the averaging step. The need to register before averaging to increase the SNR of the data is termed case 1 in this work.

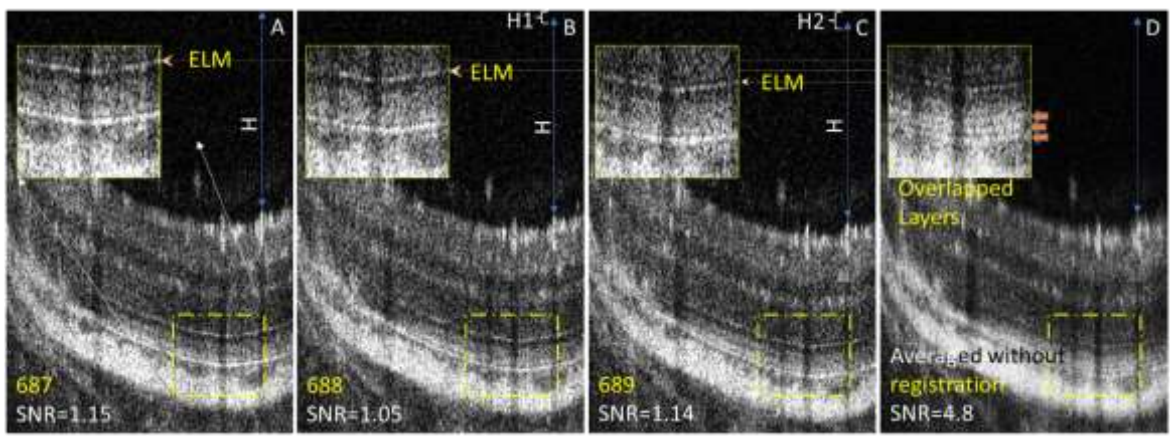

Figure 1: Image registration benefits and issues with inaccuracies: A), B), and C) show three successive B Scans having a slight decrease in height from the top of the time, with respective SNR and zoomed part showing ELM and RPE in the inset; D) is averaged B Scan of A), B) and C) without registration showing that SNR is improved but fake retinal layers appear.

#### 1.1.2. Batch processing and common dispersion values

An alternate possibility that may create fake layers is discussed in fig. 2. Raw data post-processing step to extract OCT and respective phase variance angiography data (OCT-A) utilizes requires dispersion values [11]. These values are either provided by the spectrometer manufacturer or can be estimated numerically as well [12]. Generally, single dispersion correction (for batch processing) is used which may or may not be optimal and thus may create hazy B Scans as shown in figs. 2(B-G). The need to register the data set in this condition is referred to as Case 2 in this work. Few works have been reported to make a correction [13].

A close comparison of figs. 2A and 2H, with figs. 2B-2H shows shown in are overlapped and unsharp/hazy retinal layers. The corresponding enface is shown in fig. 2J and its digital zoom section (in fig. 2I) show the overall effect. The orange-coloured horizontal arrow-shaped markers are used to highlight the existence of horizontal regions entirely giving the wrong output. Figure 2I highlights the fact that due to movement the blood vessel is wrongly depicted as broken (between 225-233).

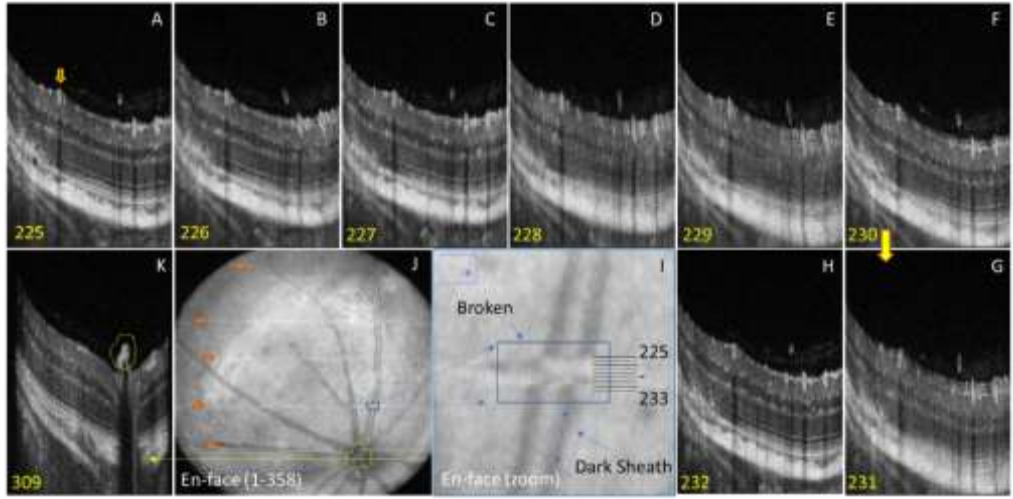

Figure 2: Effect of suboptimal dispersion correction parameters, figs. 1 (A) – (I) shows the effect of mouse eye movement as retinal layer depiction sharpness gets affected starting from (B) and ending at (H).

A general approach is to blacken out the affected region instead of giving false information if time-averaged multiple B Scan fails to remove this effect. It creates smooth-looking images with low contrast Alternatively, images are presented as it is for the user to comprehend. The worst-case scenario expects one to discard the valuable data together sacrificing animal imaging time on stage.

Interpolation techniques may subdue the effect but might add false information. Time averaging also loses the temporal resolution which may not be acceptable if OCT-A imaging is intended for flow measurements.

### 1.1.2. Shadows underneath:

Another issue is the existence of dark shadow (dark vertical column) due to existing thick blood vessels in or entering the Inner limiting membrane (ILM) in all images (shown in fig. 2(a) using a vertical orange arrow marker). The enface image (the averaged full stack of all the B Scans taken in single shot measurement) especially shows thick vessels belonging to ILM only with biased contrast with dark sheath (marked in fig. 2(I)). This issue particularly exists in OCT data. The sub-retinal fluid, floaters, vignetting, and cataracts create shadow from the top. Had these shadows were not existing, the overall enface image would have been appearing differently. The effect is more pronounced (showing a wide dark gap) under the optical cord region as shown in fig. 2(k).

If the segmentation algorithm is not robust, the effect may involve an error as these shadows might generate discontinuities in layers below them when binarized. Phase variance images analogous to figure 2 are shown in supplementary figure S1.

Several works are reported using clinical/human OCT imaging data for shadow detection and segmentation[14,15]. The correction part to remove the shadows in OCT/OCT-A data is rarely suggested using the idea of brightness matching [16,17]. Slab subtraction is shown to remove projection artifacts found to disrupt vessel continuity [18]. The best-proven method using clinical datasets so far is the Projection-resolved OCTA technique that suppresses the projection artifacts/shadows under retinal vessels of small diameter but fails to resolve IS/OS layer [19,20]. Phase variance OCT (OCT-A) enface clinical data is used to show the performance.

Standard AI-based classical techniques (for example Support Vector Machine, optical flow, and graph methods) are deployed for clinical applications as far as automated post-processing is a concern[21–23]. Deep Learning based models require thorough testing as far as the sensitivity of hyperparameters with respect to imaging data is required for optimal performance [24].

### 1.2. Motivation:

Some of the standard classical approaches for registration are all integrated into ImageJ and available in python libraries in the open domain. These, however, may not work on every dataset. Both issues (a) registration and (b) shadow identification and suppression may sometime create layer linkage problems in single-segmented layers requiring interpolating between broken or missing lines [14,25].

Available post-processing techniques are data dependent, mostly developed for human OCT data (due to ease of availability) or require human input to tweak the performance.

This work tests a simple-to-implement height adjustment technique, three Keypoint detection techniques, a hybrid model of the conventional CNN model, and optical flow for registration. To the best of our knowledge, these four methods have never been employed for the same purpose in literature. Homography or perspective transformation is used to align the images taken from visual spectrum cameras but not used to register the OCT data set [26]. The optical flow method is only used to estimate micron-scale fluid flow velocities using non-medical imaging OCT data sets [27].

The work also proposes a shadow suppressing (not detecting) technique using simple to implement analytical expression. Before that, it briefly presents a performance evaluation of five other alternatives.

The objectives are to improve the accuracy of retinal layer segmentation by reducing the above-mentioned issues in forthcoming sections.

## 2. Materials and Methods

### 2.1. Animal Husbandry and handling during the imaging:

Mice (Balb/C, 10 male and 10 female, both) are kept in the Institute Animal House and brought into the imaging facility only during the imaging sessions for a day or two. The mouse is kept under anaesthesia (isoflurane mixed with 2% oxygen). An Isoflurane vaporizer and oxygen pump are used to create the mixture. The Mouse eye is dilated by applying tropicamide and phenylephrine drops for two minutes or so. During the imaging session, an artificial tear (Gel Tear) is used to keep the cornea moist as the mouse stops blinking, naturally.

### 2.2. Post-processing steps for registration:

In our approach, a single dataset undergoes into registration process twice. The first registration is done using multiple reference frames. The algorithm offers two options to estimate the indices of these frames: (a) it allows the user to review the images and expects the index of several reference frames as input hoping that the user will enter a reference frame in the neighbourhood of the respective images those need registration or (b) it automatically estimates those indices by comparing the threshold depreciation (h2-h1>4 pixels) in ILM height in successive 5 images. These inputs are then used to perform registration to images of those indices. Afterward, a second registration is performed just using the central index of the dataset for all images.

The data set affected due to the Case 1 condition can be simply registered just to elevate the height of retinal layers by h1 and h2 pixels in successive images (in case 3 images of the same location are saved) as shown in fig.1 before averaging. The first step, however, needs estimation of h1 and h2 which is only possible if ILM is accurately segmented. This is relatively easy to perform. Tracing ILM from the side of vitreous humor (appears dark in image) a sharp gradient facilitates clear peak to gaussian fit and extracts its index. This height adjustment concept using the h1 and h2 extracted from OCT data can also be applied directly to corresponding OCT-A data.

To resolve Case 2 conditions, however, Keypoint detection techniques, namely: Scale Invariant Feature Transform (SIFT), OpenCV libraries of Oriented fast and Rotated Brief (ORB), Boosted Efficient Binary Local Image Descriptor (BEBLID) are proposed in this work [28]. These algorithms detect typically the focal points that catch the eye (for similarities to perceive a change in height) and are the areas of interest that remain constant throughout the image's change, aiding in the preservation of the crucial details while transforming. Descriptors or histograms of the photo gradients describe how these key points appear and are used to compare the keypoints in the sample image to the keypoints in the reference image. Homography 3x3 matrix containing information about the rotation and translation transformation of the target image w.r.t. the reference image is built. These keypoint detection methods are described now.

Scale Invariant Feature Transform (SIFT) is applied using the following:

1. Gaussian Blurring is applied to reduce the noise and,
2. Features are enhanced by taking the difference of Gaussians (DoG) technique,
3. Local maxima and minima values are used to remove low contrast points to detect key points,
4. Magnitude and orientation are calculated at each pixel for generating descriptors for each keypoints (128-bit vectors).

Orient Fast and Rotate Brief (ORB) applies the following algorithm:

1. Apply FAST (Features from Accelerated Segment Test) to detect features from the images,
2. Uses rotated BRIEF to calculate the descriptors of these keypoints. The rotation of BRIEF is in accordance with the orientation of the keypoints.

The Pseudo code for FAST is given as follows:

1. Select a pixel p in the image and let its intensity be $I\_p$,
2. Select a circle around it of the mask of x number of pixels,
3. Select a threshold value t and a hyperparameter n.
4. A pixel p is said to be a feature if there are n contiguous pixels in the circle, which are either brighter than $I\_(p+1)$ or darker than $I\_(p-t)$.

The Pseudo code for BRIEF is given as follows:

1. A Gaussian kernel is used to smoothen the given image,
2. Choose n location pairs of (x,y) and compare intensities at x and y. It generates a binary string of whether p(x)>p(y) or p(y)>=p(x).
3. This binary string (of 128 to 512-bit length) acts as the descriptor for a keypoint.

Homography transformation is utilized with a neural network. The idea is simple: train a neural network to find the vertices of a given reference and target image and use these points to calculate the Homography in between and transform them to register with the reference frame using those four points as anchors. It is termed here as the bounding box ML approach. Generally, a bounding box is rectangular, and one only needs four outputs from a neural network to create them, which are coordinates of reference point x, y height h, and width w. OCT retinal structure cannot be precisely bound by using rectangular bounding boxes. Thus, here polygon bounding boxes that exactly can trace can also be tested as model 2. Model architecture is a regular CNN, and the output layer will be a fully connected layer having 8 nodes, giving 8 values of coordinates. To improve the accuracy, gradient analysis is used, i.e., a hand-annotated polygon box is approximated into a quadrilateral where sharp gradient changes occur. This neural network is trained to create non-rectangular bounding boxes, giving us four pairs of coordinates, i.e., (x1, y1), (x2, y2), (x3, y3), and (x4, y4).

## 2.3. Shadow detection:

Shadows shown in fig. 2 can be considered just less bright vertical columns, with the requirement of the information needs to preserve within. The case is shown in fig. 2(K), however, also expects to generate the missing information. In some cases, it may be just conveniently extending the retinal layer maps from both sides using interpolation. Detail of classical approaches such as smearing, balancing HSV values, convolving filters, etc., are tested and directly discussed in the results section. The detection of the shadow is considered a segmentation problem. Standard U-Net and RCNN are tested [31,32].

### 2.3.3. Method to update the Pixel Value under shadows:

The linear function is shown in eq. 1 is used on 8-bit data to update the pixel values inside shadow regions once their coordinates are estimated. The n is the number of bit resolution of data. In our case, to optimize the speed of processing we have converted the data into 8 bits.

$$Pix\_val_{New} = \alpha \times (2^{n-1} - Pix\_val_{old})/2^{n-1} \qquad (1)$$

## 2.4. AI and layer thickness estimation:

After the images are stabilized U-net is used to segment out the pixels at retinal boundaries. Cycle GAN is also used for this aim, but the method is discarded as a preferred choice due to poor results. Manually annotated layer maps are used as input to train the U-net model as described earlier. Keras (TensorFlow) framework is used. The model optimizes the combination of Dice and Binary Cross Entropy loss.

# 3. Results

## 3.1. Registration: Classical, AI, and Hybrid

### 3.1.1. Homography transformation approaches

SIFT is applied to image fig 3(A) using the reference image. The estimated keypoints are shown in fig. 3(B) after registration is done. The red circles are keypoints while the green lines represent good matches. It clearly shows that this technique is not estimating sufficient keypoints (green lines). For ORB, 16 pixels are used for creating a mask in FAST. The figs. 3(C) shows that using ORB relatively more keypoints are incorrectly matched than in SIFT. The hybrid of ORB and BEBLID is tested by applying the contrast boosting step. The results are shown in fig. 3(D) are better. In conclusion, these methods may work with images that have sharp edges and highly contrasting features to detect keypoints but fails for relatively warped cases.

The optical Flow method (from sci-kit-image libraries) is used in the next stage. In this method, the algorithm computes optical flow for each pixel, it calculates the vector (u,v) for every pixel such that reference(x,y) = target(x+u,y+v). It can be then used for image warping and transformation. The results, shown in fig. 3(E) are not better than ORB + BEBLID approach. Correlation values estimated for SIFT, ORB, ORB+BEBLID, and Optical flow are 0.396, 0.54, 0.598, and 0.535, respectively. The results of these methods are not encouraging when applied on a stack. The average correlation value estimated on 4 different stacks each consisting of 360 images always remained less than 0.6.

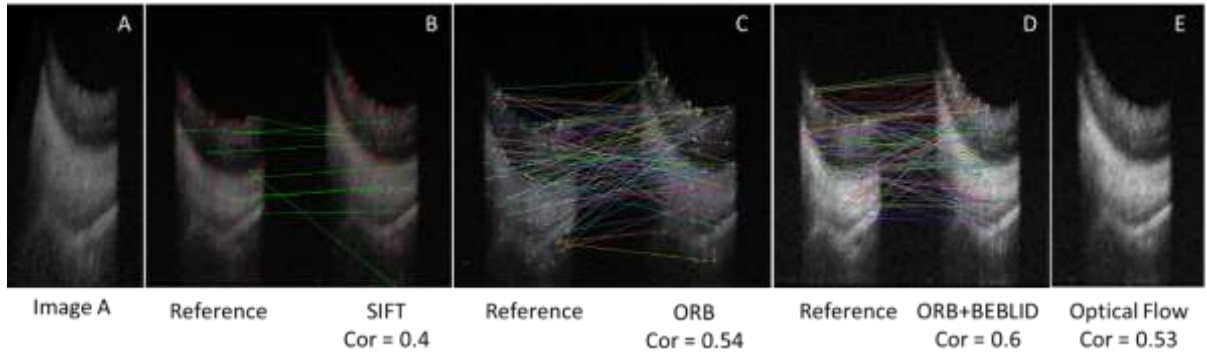

Figure 3: Output of algorithms for image registration. A) sample image A is the image that needs registration, B), C) and D) shows the keypoints obtained from mapped between reference image and sample image (after registration), obtained by SIFT, ORB and combination of ORB and BEBILD, respectively. E) shows the registered image using Optical Flow. The correlation value remains below 0.6 for all.

### 3.2. AI and Optical flow hybrid:

Four hundred images are annotated for creating the training dataset using a hybrid of model 1 in the shape of a trapezium. For model 2, polygon segmentation requires too much computing power and has a large time complexity, when compared for practical use purposes, as the number of sides of polygon increases, the complexity increased as well. These annotation models 1 and 2 are shown in fig 4(A). The images were annotated in State-of-The-Art COCO-JSON format, a specific JSON structure dictating how labels and metadata are saved for an image dataset. This format is used the most for object detection/image segmentation tasks. Annotations were done using Make Sense software.

The neural network model is a deep CNN with 245M parameters, with layer details depicted in fig 4(B). Its initial parameters are listed in fig. 4(C).

Different filter sizes 7x7, 5x5 and 3x3 filters have been used by keeping in mind that one needs to consecutively check for smaller and smaller features. Initial loss (using MSE as the loss function) came high ($3\times10^5$). It is observed that the dataset has an average mean and standard deviation of 32.9 and 36.06, respectively. The target vector is normalized using inbuilt functions of PyTorch and the addition of this normalization is done to the output vector of our model, thus normalizing the two vectors needed to calculate the loss. It reduced the loss value to 14.3689 but this level still is unacceptable as the correlation value is 0.627. The model is further modified to incorporate another feature of CNNs referred to as Batch Normalization. Three layers of Batch Norm (bn1, bn2, and bn3) are added after each Max Pool layer. This resulted in much better results than before, but still, too high for normalized inputs and outputs, as those were normalized to be centered around 0. Initially, the model had no normalization, hence the loss was very high for an object detection task. The final training and testing loss, in fig. 4(D), shows that 10 epochs are more than sufficient to achieve desired loss value. The method is tested on the sample image shown in fig. 4(F) using reference image 4(E). Its registered image is shown in fig. 4(G). The correlation value for this method using a sample is 0.71. The average correlation value for 4 stacks is 0.7056.

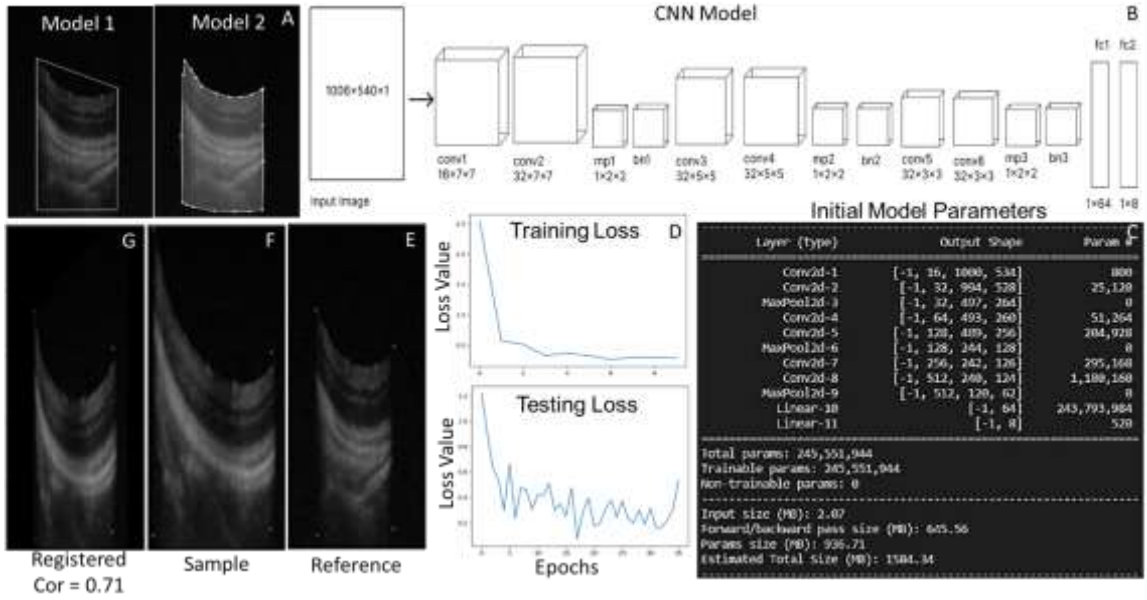

Figure 4: Registration using deep CNN; A) shows annotation models, B) in-house developed CNN model architecture, C) set model parameter details are given, D) Loss value vs epochs, E), F) and G) are reference, sample, and registered images.

### 3.3. Suppression of shadows:
### 3.3.1. Classical approaches for shadow detection:
Classically, a 3 x 3 kernel can be convolved for vertical edge detection as a first step. A 3 x 3 kernel once convolved with the sample image (fig. 5(A)) although detects vertical edges also detects several other vertical structures (which may be present due to noise) shown in fig. 5(B).

Another approach is simply smearing the whole image column-wise. These dark columns of shadows are generally less than thin so axis-0 rolling averaging (fig. 5(C)) with a window having a width of n=10 pixels is tested but it smears in the necessary information losing the contrast resolution. Tweaking Hue-saturation-value as H-S controls the color and value controls the brightness may help. A linear function (eq. 1) with alpha value = 1.3 is tested on a value that controls the dark spots while keeping the light spots as it is. The results (fig. 5(D)) demonstrate that the variation in brightness of the shadow and non-shadow part has been reduced but still we can see that the shadows have not been removed.

The information of adjacent columns is exploited assuming they have a spatial similarity. It may locate the shadows and give the corresponding brightness difference. Equation 2 is used; it subtracts the brightness of each pixel from the previous pixel in the same row. Results do contain some information about the shadow columns but are unsatisfactory as the images (fig. 5(E)) have granular (noisy) and not well-defined boundaries.

$$Img_{((i,j))} = Img_{((i,j))} - Img_{((i,j-1))} \qquad (2)$$

Finally, blurring and then thresholding are also tested. This method is often used in computer vision tasks to detect edges by first averaging/blurring the image with a 3x3 filter and then thresholding the difference for a particular value. The values less than the threshold value are given a value of 0 and more than the threshold are kept the same. This method also doesn't work as there is a lot of noise in the images, and a lot of random bright/dark pixels all over the image (fig. 5(F)).

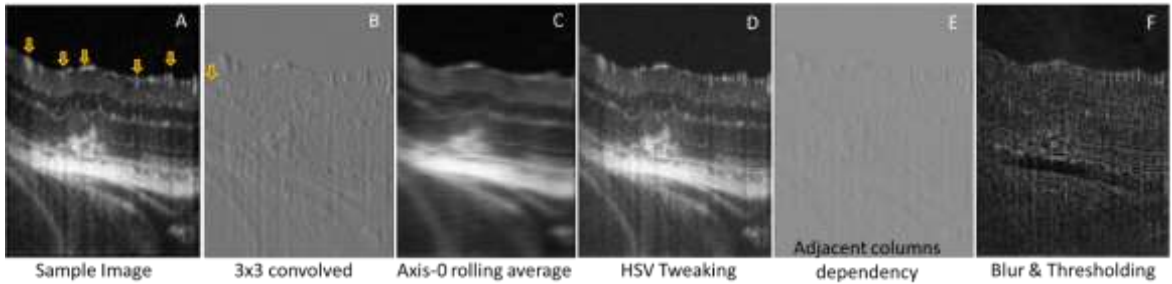

Figure 5: Classical approaches for Shadow removal.

### 3.3.2. AI models for shadow detection:

Unlike classical approaches, which affect the whole image, an AI-based problem first identifies the region containing shadow in the first step and expects normalizing the patches afterward. The detection of the shadow is considered a segmentation problem. Computer Vision Annotation Tool (CVAT) is used to prepare a labeled dataset using 359 images. These images are registered, first. The bounding box annotation approach is used as shadows were mostly columns and can be easily segmented as a rectangular box. Codes are run till 60 epochs for saturation behaviour. Binary masks are extracted from the labeled dataset. All pixels in the shadow region are given a value of 1 and other pixels were given a value of 0. The OCT image data are used as input to the U-net model and the corresponding masks are used as a target. U-net model on the labeled dataset to detect and predict shadows. U-net is trained from scratch by using a combination of dice and focal loss as a loss function, with a batch size of 2 and a learning rate of 0.0001.

Results from U-Net are shown in fig. 6(D) are less encouraging with a loss value of 0.76. One reason may be as it is a semantic segmentation method it treats all the shadows (in the sample image) as a single object and aims to segment them all from the image. We note that the number of shadows, their width, and their height is not the same in various sample images.

R-CNN, an instance segmentation approach, is tested further. Instance segmentation treats all the shadows as separate objects and is used for multi-object detection tasks. FastRCNNPredictor which is a predefined model provided by PyTorch is used. Faster RCNN is based on the Resnet50 backbone. This model requires all the shadows, separately annotated as individual entities so one more time annotation is performed. ResNet50 backbone with a learning rate of 0.00001 for 5000 epochs is trained. Out of the predicted regions of shadows, only those regions with a confidence score above 0.8 are accepted and, in those regions the pixels with a value equal to or greater than 0.5 are given a value of 1. The loss value of 0.903 is achieved with this model.

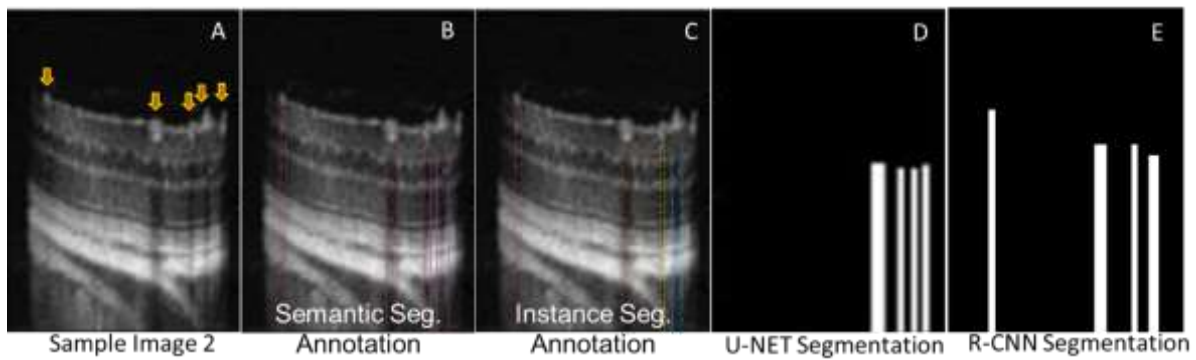

Figure 6: AI-based segmentation of region under the shadows.

### 3.3.3. Pixel Value Manipulation:

The correlation value between data with and without registration for this data set is 0.703. Once the location of shadows is found, the next step requires changing the brightness/pixel grey values. Equation 2 is used to manipulate the pixel value under the shadow only. Sensitivity analysis (simple brute force approach) is performed to find the optimized (visual perception is used) value of α=1.3. The final shadow removed registered image (AI adjusted) image and respective original images are shown in fig. 7. Figures 7(A), 7(B), and 7(G) are enface and B Scans (index 316 and index 341) of original data (stack of 360 B Scans) and figures 7(F) 7(C) and 7(H) are corresponding AI segmented pixel value adjusted images for shadow suppression. The effect is visible by comparing fig. 7(B) with fig. 7(C) and 7(G) with 7(H). Shadows, however, are not entirely removed but suppressed. Corresponding normalized differences (with corresponding shadow locations) are also shown in Fig. 7(D) and 7(I), basically equivalent to the mask created by AI to segment these locations. Figures 7(B) and 7(E) show that the chosen value of alpha is able to suppress the existence of relatively thin shadows (highlighted with arrow marker with light blue boundary and white inner core) only. The biggest possible dark region right below the optical cord is adjusted as well, however, interpolation to the missing retinal layer maps is not included in this work. The difference between figs. 7(B) and 7(E) is shown in fig. 7(G) depicting the overall changes made by AI in this B Scan. The major effect of suppressing the shadow is observed in the enface of the full stack of 360 B Scan. The enface of the original images (Fig. 7(A)) has suffered the case 2 registration issue and shows dark ILM vessels, prominently. The AI-adjusted enface however has only a few jitters and dark vessels after post-processing. The histogram data obtained for the respective full stack is shown in figs. 7(E) and 7(G). The difference in the mean pixel value of shadow-adjusted data increased (1%) from 51.77 to 52.29 and the standard deviation decreased (1.29%) from 55.25 to 54.54. The number of zeros (mode value) decreased (77%) from $2.57 \times 10^7$ to $5.73 \times 10^6$. It is reflected in the increased smoothness of the histogram curve. The data set using several values of α is found to be having poor performance as far as loss in standard deviation, zero-valued pixels, and increment in mean pixel value is concerned. We found that the positive mean pixel values, negative standard deviation, or the number of zeros can be used as optimization factors to estimate the value of optimal α rather than using brute force sensitivity analysis.

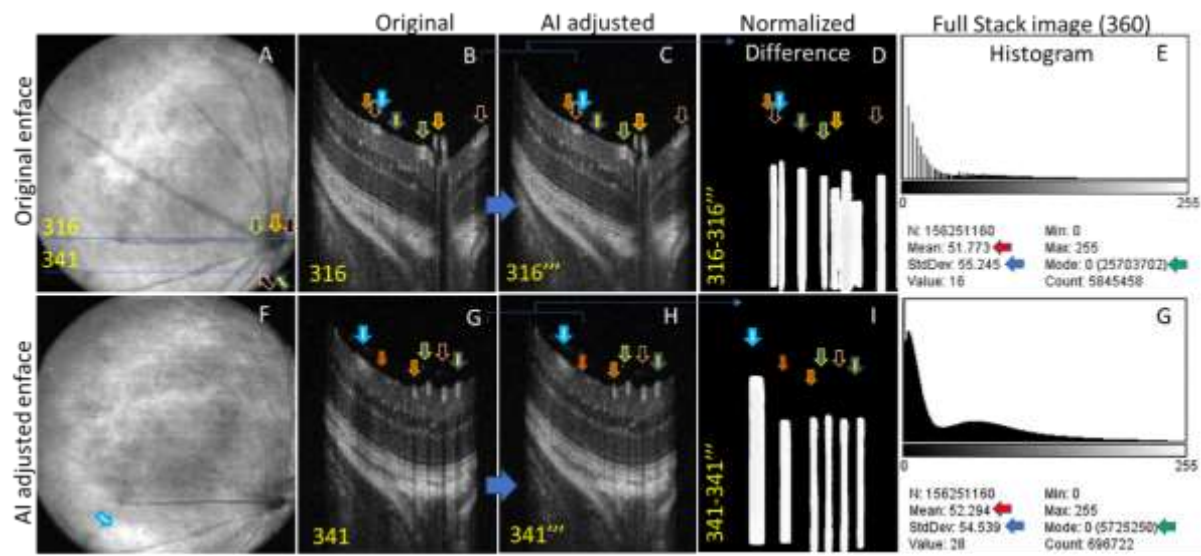

Figure 7: AI segmented pixel value adjusted images for shadow suppression, A) enface image without shadow suppression, B) and C) are B Scans showing the existence of shadow region, D), E) and F) are corresponding enface and B Scans image after AI-based shadow suppression, G) is difference between images shown in C) and F).

### 3.4. AI-based retinal layer thickness estimation:

The flow chart of the algorithm is shown in supplementary figure S2. The implementation of U-Net and associated parameters are explained elsewhere [24]. The in-vivo imaging session is performed on twenty mice. The OCT B Scans and OCT-A enface images are post-processed. Figure 8 shows the retinal layer thickness automatically estimated using an AI-based app. BALB/c female mouse data is used to verify the difference with published data. Enface OCT and its phase variance images are shown in Fig. 8(A) and (B). B Scan (for example) is shown in fig. 8C. Its zoom version in fig. 8(D) is used to depict the respective retinal layers along with pixels and lengths in micrometers. It appears that the AI-based data is closer to manually segmented reported data than other tools for this B Scan[33]. The supplementary figure S3 comparatively depicts the binary mask created using U-net, AI-Pipeline and justifies the improvement made.

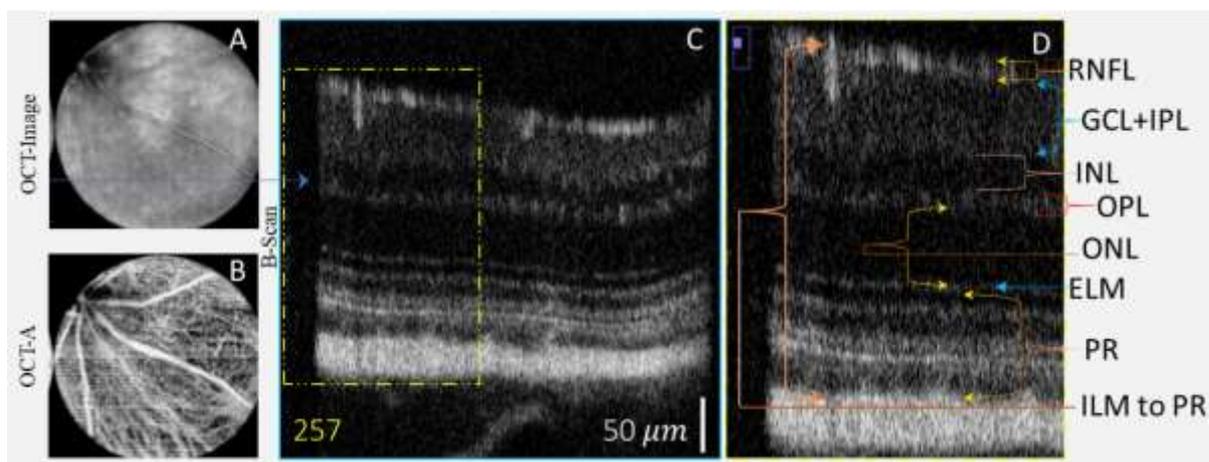

Figure 8: AI-based Automatic Retinal Layer Thickness Estimation of BALBc female mouse; A) and B) are Enface OCT and Phase-Variance images, C) shows B Scan with 257 indices (marked in OCT image with blue line), and D) is just the zoom section from an earlier image with retinal layers/sections segmented.

The detail is given in Table 1. The reported thickness from ILM till the end of PR is 209 mm. Gender is not mentioned in this work. AI segmented average distance (for all B Scans using more than 20 mice) in pixels is found to be 251 pixels. It indicates that the axial resolution of the system is 0.836 microns per pixel. The overall estimation has 6% of error.

Table 1: Retinal thickness estimation comparison for BalBc female

| Layer | Pixels | Thickness estimated using AI | Thickness reported[33] |
|---|---|---|---|
| RNFL | 20 | 16.72 | 19.32 |
| GCL+IPL | 53 | 44.31 | 45.09 |
| INL + OPL | 25+21 | 20.9+17.5 = 38.4 | 41.92 |
| ONL | 52 | 43.47 | 46.09 |
| ELM | 12 | 10.03 | |
| PR | 81 | 67.72 | 59.86 |
| ILM to PR | 251 | 209.20 | 209.20 |

### 4. Conclusions

OCT and OCT-A images are obtained from a compact table-top multi-spectral imaging system that is developed in-house. The work presents AI-based post-processing add-ons required for accurate retinal layer segmentation, especially the suppression of shadow beneath the thicker ILM structures. It is shown that the classical keypoint detection methods, conventional neural network, and AI model for medical imaging underperform for the speckle noise-ridden mouse eye OCT data when it comes to segmenting low contrast regions. Tweaking the AI model by inserting the batch normalization process has provided an acceptable loss value. The retinal thickness estimation accuracy is 94% when 359 images are used for training. The method has limitations as performance severely depends on variation available in the training dataset. A healthy animal has relatively fewer morphological variabilities in OCT data than an animal suffering from any disease. Segmented data with and without AI post-processing has a difference as without post-processing gaps in single retinal topography affect the overall estimation. Post-processing is added as a prior step in the layer segmentation pipeline. The customized code is converted into a user-friendly app that allows the user to add his/her annotated data set for training purposes.

Funding: DST-SERB: IMPRINT-2, IMP/2018/001045, CDRC/22-23/GR1/P10/03.

Institutional Review Board Statement: The animal imaging sessions are performed under approved protocols BT/IEAEC/2018/2/03 compliant to the CPCSEA, Govt of India and the approval of the Institute Animal Ethical Committee at IIT Roorkee.

Data Availability Statement: Data will be provided on request.

Acknowledgments: Author would like to thank Netra Systems, Inc. for assisting in development the combined animal imaging facility. Author also acknowledges Mr. Pranjal Minocha and Mr. Simardeep Singh, EPH, Physics IITRs' help for Python library integrations to convert codes into app.

Conflicts of Interest: The author declares no conflict of interest.

# Supplementary

## S1. Effect of Eye movement and requirement of Registration

OCT-A Data showing the effect of mouse eye movement when optimal dispersion coefficients for index 225 are used on index 227-231.

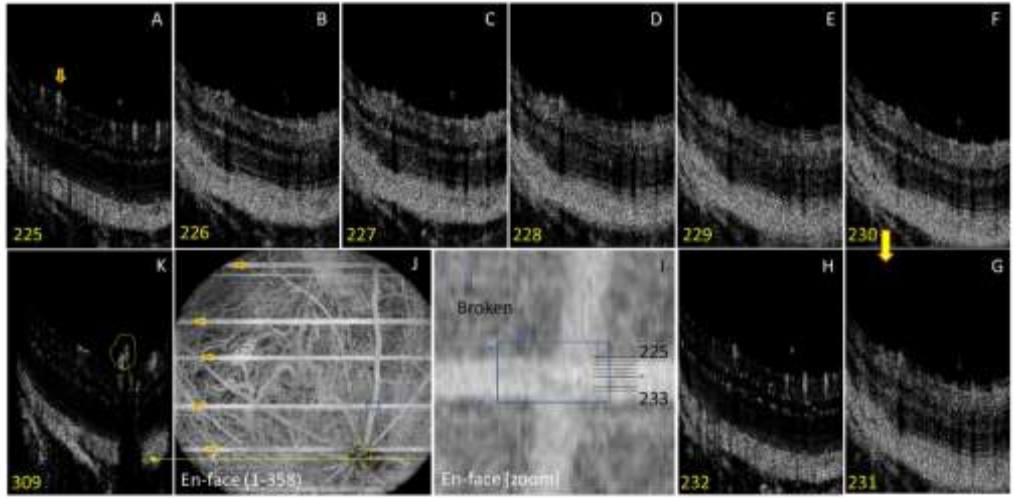

Figure S1: Effect of Mouse eye movement, figs. 1 (A) – (I) shows effect of mouse eye movement as retinal layer depiction sharpness gets effected starting from B and ending at H.

## S2. AI Pipeline

Figure S2 illustrates the flow of the algorithm.

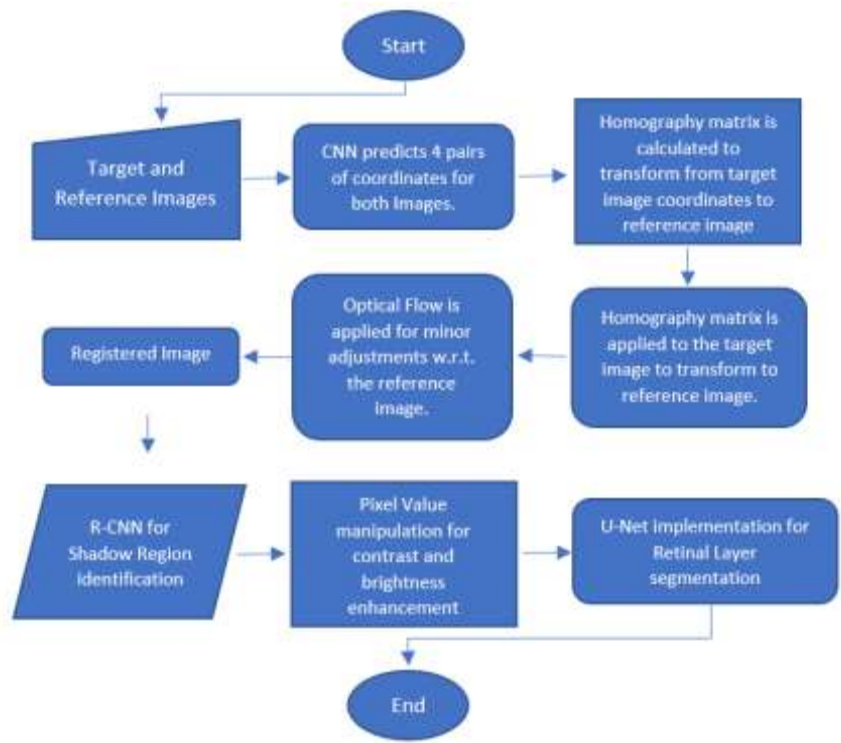

Figure S2: Flow Chart.

## S3. Retinal Segmentation with and without AI Pipeline

Without accurate segmentation standard AI model (U-net) shows presence of broken retinal layer under the ILM blood vessels causing shadowed columns. AI Pipeline (that first suppress their presence by elevating the pixel value brightness) performs relatively better.

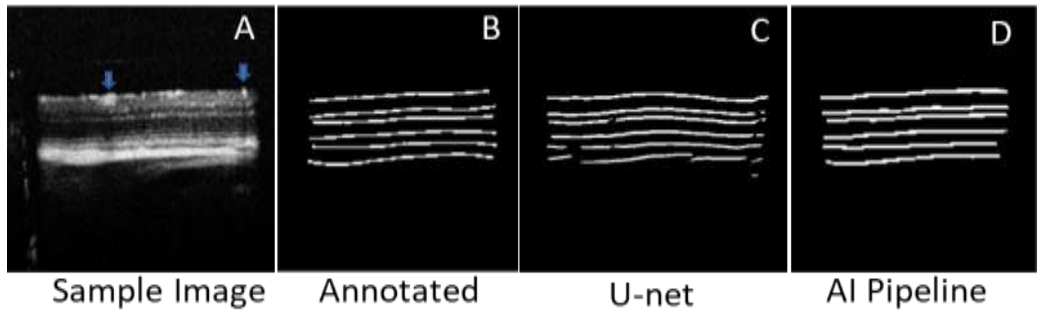

Figure S3: Retinal Segmentation with and without AI Pipeline. A) shows the sample image, B) manually annotated binary mask, C) Mask created by U-net model without removal of shadows and registration, D) Binary Mask created using AI-Pipeline.